\documentstyle [a4,12pt]{article}
\begin{document}

\title{Vortices On ``Rail Road Track'':\\
A Possible Realization of Random Matrix Ensemble}

\author{Yang  Chen and Henrik Jeldtoft Jensen \\
        Department of Mathematics\\
         Imperial College\\
        180 Queen's Gate, London SW7 2BZ, U. K.}
\maketitle

\begin{abstract}
The thermodynamics of interacting vortices constricted to move in
one-dimensional tracks carved out on superconducting
films in the extreme type-II limit is mapped into the Coulomb gas model
of random matrices. Application of the Selberg Integral
supply {\it exact} expressions for the constitutive relation
for two configurations of tracks.
\end{abstract}
\par
\noindent To appear in J. Phys. Cond. Matter.
\newpage
In the present paper we derive an {\it exact} expression for the constitutive
relation between the external magnetic field, $H$, and the magnetic
induction, $B$, inside a type II superconductor, i.e., we determine the
relationship $B=B(H)$. The induction inside the superconductor
is determined by the density, $n$, of magnetic vortices. Each vortex
carries one flux quantum $\Phi_0$, hence $B=\Phi_0 n$.\cite{tinkham}
In order to be able to find an exact expression for $B(H)$ we restrict
ourselves to thin superconducting films. The interaction energy
of two vortices depends in this case logarithmically on the
distance of separation of the two vortices. Furthermore, we assume
that the vortices are confined to move on tracks engraved into the
superconducting film, see Fig. 1. This experimental situation has
in fact been studied by Pruymboom et al.\cite{kes} The thickness
of the film is reduced in the track. Since there is a loss of
condensation energy in the core of a vortex, it will be energetically
favourable to position a vortex with the core on the track rather
than outside the track, where the core would have to drill its
way through a thicker layer of superfluid.

The combined simplification of logarithmic interaction and restriction
on the configurational degrees of freedom allow us to map the problem
onto the Coulomb gas model of random matrices.\cite{dyson} The
partition function is known exactly for this model and the constitutive
relation readily derived.

Let us now describe the derivation of the relation between $H$ and
$B$. The partition function for a fixed number of $N$ two
dimensional vortices on a track is given by:
\begin{equation}
Z_N=\left(\prod_{a=1}^{N}\int_0^L
{dx_a\over l}\right)
{\rm exp}\left(-\beta E[x_1,\ldots ,x_N]\right),
\end{equation}
where
\begin{equation}
E[x_1,\ldots,x_N]=NE_c-E_1\sum_{1\leq a<b\leq N}{\rm ln}
\left({|x_a-x_b|\over \lambda_{\rm eff}}\right),
\end{equation}
where $E_c(>0)$ is the energy of the vortex core and $E_1(>0)$
sets the scale of the vortex interaction.\cite{minnhagen-rev}
Both $E_c$ and $E_1$ are temperature dependent.\cite{minnhagen-rev}
The length of a the tracks are denoted by $L$, see Fig. 1.
The scale of the logarithmic
interaction  is set by the effective magnetic penetration depth,
$\lambda_{\rm eff}$, of the film. Expressed in terms of the bulk
penetration depth $\lambda$ and the film thickness $d$ one has
$\lambda_{\rm eff}=\lambda^2/2d$.\cite{pearl}
$\beta=1/T$ is the inverse temperature. $N$ is related to the magnetic
induction $B$ through $BLR=N\Phi_0$, where $R$ is the distance between the
individual tracks, see Fig. 1.
$l$ the length scale of the configuration space and may be
taken to be equal to $\xi(0)$, the coherence length at zero temperature.
We assume that $L< \lambda_{\rm eff}$ and  $R\gg \lambda_{\rm eff}$.

The condition $\lambda_{\rm {eff}}\gg L$
ensures that the logarithmic interaction
is valid for all possible separations of a pair of vortices
on the same track. Beyond the distance $\lambda_{\rm eff}$ the
vortex interaction decays exponentially due to the diamagnetic
screening currents.\cite{tinkham,pearl} Consequently, we can neglect
the interaction between vortices on different tracks if the condition
$R\gg\lambda_{\rm eff}$ is fulfilled.

It should be noticed that
the vortices will stay in the tracks for  magnetic fields less
that a certain filed strength  $H^*$. At $H^*$ the repulsion
between the flux lines becomes so strong that it will be
able to over come the pinning force exerted by the carved
track. The actual value of $H^*$ will depend on the depth of
the groove and in a first approximation will be equal to
the $H_{c_1}$ of the bulk material. The partition function in
Eq. 1 applies for fields $H<H^*$.

The partition function can be re-written as:
\begin{equation}
Z_N={\rm e}^{-N{\bar {E_c}}}\left({L\over l}\right)^N
\left({L\over \lambda_{\rm eff}}\right)^{N(N-1){\bar {E_1}}}
\left(\prod_{a=1}^{N}\int_0^1dt_a\right)
\prod_{1\leq a<b\leq N}|t_a-t_b|^{2{\bar {E_1}}},
\label{partion}
\end{equation}
where we have introduced the following dimensionless energies:
\begin{equation}
{\bar {E_c}}={E_c\over T},\quad\quad\quad {\bar {E_1}}={E_1\over 2T},
\end{equation}
and the dimensionless integration variables $t_a$.
The multiple integral in Eq. \ref{partion} is a special case of the
Selberg integral,\cite{selberg} with its aid we obtain the partition
function and the free energy of this system:
$$
-{F_N\over T}=\ln Z_N
=N\left[{\rm ln}(L/l)-{\rm ln}\Gamma(1+{\bar {E_1}})-
{\bar {E_c}}\right]+N(N-1){\bar {E_1}}{\rm ln}(L/\lambda_{\rm eff})
$$
\begin{equation}
+\sum_{j=0}^{N-1}{\rm ln}\left[{\Gamma(1+(j+1){\bar {E_1}})
\Gamma^2(1+j{\bar {E_1}})
\over \Gamma(2+ (N-1+j){\bar {E_1}})}\right].
\end{equation}

An expression for the lower critical field $H_{c_1}$ is obtained
from the condition that the Gibbs free energy of the state
with a single vortex $F_1-H_{c_1}BV/4\pi$ ($V=LRd$ is the volume
per track) is equal to the Gibbs
free energy of the state without any vortices
$F_0=0$.\cite{tinkham} The result is
\begin{equation}
H_{c_1}= {4\pi\over\Phi_0d}(E_c-T\ln(L/l))
\end{equation}

The constitutive relation is obtained in the following way
which makes use of the fact that $H$ acts as the chemical
potential for the vortices and $\frac{dG}{dN}=0$, where $G$ is the Gibbs
free energy for the system:\cite{degennes}
\begin{equation}
\frac{\partial F_N/T}{\partial N}={1\over4\pi}{\Phi_0\over LR}{H\over T}V
\end{equation}
We simplify the above expression in the large $N$ limit, by
replacing the sum of the logarithms by an integral while discarding the
differences between $N$ and $N\pm 1.$ With the aid
of $
\Gamma(ax+b)\approx (2\pi)^{1/2}{\rm e}^{-ax}\left(ax\right)^{ax+b-1/2},
\;a>0,\;b>0,\;x\gg 1$ we find
\begin{equation}
\frac{dH}{4\pi T}\Phi_o=2X\ln(4\lambda_{\rm {eff}}/L)+\ln X,
\end{equation}
where $X:=N{\bar E}_1=(BLR/\Phi_o){\bar E}_1,$ and we have discarded the
$N$ independent and $O(N^{-1})$ terms.
The magnetic permeability $\mu$ in the high
field limit, can be found by further discarding the ${\rm ln}X$ term
\begin{equation}
B=\mu_{o}H,
\end{equation}
with
\begin{equation}
\mu_{o}=\frac{d\Phi_o^2}{4\pi E_1LR\ln(4\lambda_{{\rm eff}}/L)}.
\end{equation}
The constitutive relation has a weak logarithmic correction for not too large
$H.$ This is found by the substitution
$X=X_o+\delta X,$
where
\begin{equation}
\frac{\delta X}{X_o}=-\ln \left(\frac{\frac{dH\Phi_o}
{8\pi T\ln(4\lambda_{{\rm eff}}/L)}}{1+\frac{dH\Phi_o}
{4\pi T}}\right),
\end{equation}
which in turns supplies a non-linear permeability
$\mu(H)=\mu_o(1+\delta X/X_o).$ The permeability is slightly reduced
from the its high field values which is to be expected.

So far we have considered the ensemble where the position of the
vortices are restricted  to a straight line of length $L$.
Let us now consider a system in which the vortices
are confined to a circular track of radius $R_o$.
In this case we avoid edge effects. The thermodynamics of the
the logarithmically interacting vortices on
a ring are obtained from Dyson's circular
ensemble.\cite{mehta}
The interaction energy of the vortices is given by
\begin{equation}
E^c[\theta_1\ldots\theta_N]=NE_c-E_1\sum_{1\leq a<b\leq N}
{\rm ln}\left({R_o\over \lambda_{\rm eff}}
|{\rm e}^{i\theta_a}-{\rm e}^{i\theta_b}|\right),
\end{equation}
and the partition function
\begin{equation}
Z^c_N=\left(\prod_{a=1}^{N}\int_{0}^{2\pi}
{R_o\over 2\pi l}d\theta_a\right){\rm exp}
\left(-\beta E^c[\theta_1\ldots \theta_N]\right),
\end{equation}
where $R_o(\ll \lambda_{\rm eff})$ is the circumference of the circle.
The constitutive relation is in this case easily obtained directly
from the difference $F_{N+1}-F_N$ in free energy. We have
$$
-\left(\frac{F^c_{N+1}-F^c_{N}}{T}\right)
$$
\begin{equation}
=
2N{\bar E}_1\ln(R_o/\lambda_{{\rm eff}})+
\ln\left[\frac{\Gamma\left((N+1){\bar E}_1\right)}
{\Gamma(N{\bar E}_1)}\right].
\end{equation}
For large $N$ and with the aid of large $x$ behaviour of $\Gamma(ax+b)$,
we find,
\begin{equation}
\frac{\Phi_o H}{8\pi^2 R_od T} V=2\ln
(\lambda_{{\rm eff}}/2\pi R_o)Y+{\bar {E_1}}{\rm ln}Y,
\end{equation}
where
\begin{equation}
Y:=N{\bar {E_1}},
\end{equation}
The result for the permeability in the high field limit
reads
\begin{equation}
\mu^c_{o}=\frac{1}{32\pi^2E_1R_o}\frac{\Phi_o^2}
{\ln(\lambda_{{\rm eff}}/2\pi R_o)}
\end{equation}
a result similar to that of the straight track.

We have derive an exact expression for the constitutive
relation for an artificially patterned superconducting thin
film. The result should be directly experimentally accessible.
Pinning along the the engraved track may in principle
influence the motion of the vortices. Although, it appears
that for deep grooves pinning should effect the motion
parallel to the track rather insignificantly.

Our future extension of the present calculation will be to
apply random matrix theory to the dynamics of the vortices
in the constricted geometry.

\newpage
\begin{center}{\bf  Captions} \end{center}

\noindent{ \bf Figure 1.} \\
Sketch of a superconducting thin film with tracks engraved into
the surface.

\vskip3cm

\noindent{\bf Figure 2.}\\
The induction $B$ as function of the external magnetic field
$H$ obtained from Eq. 8 for the following set of parameters:
$L = 1 \mu m = R/10$, $d=10$ \AA, and $\lambda = 1500$ \AA.
The temperature is supposed to be 10K.

\end {document}